\documentclass[10pt, conference, compsocconf]{IEEEtran}
\ifCLASSINFOpdf
\else
\fi
\usepackage[binary-units=true,group-separator={,}]{siunitx}
\usepackage{latexsym}
\usepackage{indentfirst}
\usepackage{times}
\usepackage[titles]{tocloft}
\usepackage[colorlinks=true,allcolors=blue,breaklinks,draft=false]{hyperref}
\usepackage{colortbl}
\usepackage{fancyhdr}
\usepackage{ifthen}
\usepackage{url}
\usepackage{amsmath}
\usepackage{amsfonts}
\usepackage{mathtools}
\usepackage[ruled,lined,linesnumbered,boxed,vlined]{algorithm2e}
\usepackage{xfrac}
\usepackage{xcolor}
\usepackage{lineno}

\usepackage{chngcntr}
\counterwithout{equation}{section}

\usepackage{caption}
\usepackage{graphicx}
\usepackage{subcaption}

\usepackage{array}
\newcolumntype{L}[1]{>{\raggedright\let\newline\\\arraybackslash\hspace{0pt}}m{#1}}
\newcolumntype{C}[1]{>{\centering\let\newline\\\arraybackslash\hspace{0pt}}m{#1}}
\newcolumntype{R}[1]{>{\raggedleft\let\newline\\\arraybackslash\hspace{0pt}}m{#1}}
\usepackage{multirow}

\newcommand\STR{\textit{Str}}
\newcommand\state[1][{}]{\ensuremath{{#1}}}
\newcommand\State[1][{}]{\ensuremath{q_{#1}}}
\newcommand\States{\ensuremath{q}}
\newcommand\IStates{\ensuremath{I}}
\newcommand\IState[1][{}]{\ensuremath{I_{#1}}}
\newcommand\StateSet[1][{}]{\ensuremath{Q_{#1}}}
\newcommand\StateSets{\ensuremath{Q}}
\newcommand\StartState{\ensuremath{\State[0]}}

\newcommand\FinalStates{\ensuremath{F}}
\newcommand\NrStates{\ensuremath{\lvert\StateSets\rvert}}

\newcommand\TF{\ensuremath{\delta}}
\newcommand\TFE{\ensuremath{\TF^*}}

\SetKwFor{WhilePar}{while}{do in parallel}{done}
\SetKwData{STATE}{state}
\usepackage{tikz}
\usepackage{tkz-graph}
\usetikzlibrary{positioning}
\usetikzlibrary{matrix}
\usetikzlibrary{shapes}
\usetikzlibrary{shapes.symbols}
\usetikzlibrary{trees}
\usetikzlibrary{calc}
\usetikzlibrary{arrows}
\usetikzlibrary{shadows}
\usetikzlibrary{chains}
\usetikzlibrary{automata}
\usetikzlibrary{decorations.pathmorphing,backgrounds,fit,positioning,chains}
\usetikzlibrary{decorations.markings}

\usepackage{listings}
\usepackage{framed,color}
\tikzset{
    state/.style={
        circle,
        draw=black, very thick,
        minimum size = 1cm,
    },
    smallstate/.style={
        circle,
        draw=black, very thick,
        minimum size = 3mm,
        inner sep=0.8mm,
    },
    smallfinalstate/.style={
        circle,
        double,
        draw=black, very thick,
        minimum size = 3mm,
        inner sep=0.8mm,
    },
    description/.style={
        rectangle,
        draw=white,
        mininum height = 2em,
        inner sep = 0pt,
    },
    tuborg/.style={
        decorate
    },
}

\usepackage{fancyhdr}
\fancyheadoffset[L]{15mm}
\fancypagestyle{firststyle}
{
	\fancyhf{}
	\fancyhead[L]{\footnotesize
	\textbf{\vspace{0mm}This work has been superseded by the following
		paper published at the 46th International Conference on Parallel
		Processing ICPP'2017: \href{http://elc.yonsei.ac.kr/publications/ICPP_SFA_preprint.pdf}{\color{black}[LINK here].}}}
}

\begin{document}
\title{Efficient Construction of Simultaneous Deterministic Finite Automata on Multicores Using Rabin Fingerprints}


\author{\IEEEauthorblockN{Minyoung Jung and Bernd Burgstaller}
\IEEEauthorblockA{Department~of Computer Science\\
Yonsei University\\
Seoul, Korea
}
\and
\IEEEauthorblockN{Johann Blieberger}
\IEEEauthorblockA{Institute~of Computer Aided Automation\\
Vienna University of Technology\\
Vienna, Austria
}
}

\maketitle
\thispagestyle{firststyle}

\begin{abstract}
The ubiquity of multicore architectures in modern computing devices requires
the adaptation of algorithms to utilize parallel resources. String pattern matching
based on finite automata (FAs) is a method that has gained widespread use 
across many areas in computer science. However, the structural properties of
the pattern matching algorithm have hampered its parallelization. 
To overcome the dependency-constraint between subsequent matching steps and exploit
parallelism, simultaneous deterministic finite automata (SFAs) have been
recently introduced.
Given an FA~$A$ with $n$~states, the corresponding SFA~$S(A)$ simulates $n$ 
parallel instances of FA~$A$. However, although an SFA facilitates parallel FA matching,
SFA construction itself is limited by the exponential state-growth problem, which
may result in $O(n^n)$ SFA states for an FA of size~$n$. The substantial
space requirements incur proportional processing steps, both of which
make sequential SFA construction intractable for all but the smallest problem sizes. 

In this paper, we propose several optimizations for the 
SFA construction algorithm, which greatly reduce the in-memory footprint and
the processing steps required to construct an SFA.
We introduce fingerprints as a space- and time-efficient way to represent SFA states. 
To compute fingerprints, we apply the Barrett reduction algorithm
and accelerate it using recent additions to the x86 
instruction set architecture.
We exploit fingerprints to introduce hashing for further optimizations.
Our parallel SFA construction algorithm is nonblocking
and utilizes instruction-level,
data-level, and task-level parallelism of coarse-, medium- and
fine-grained granularity. We adapt static workload distributions and 
align the SFA data-structures with the constraints of multicore memory hierarchies, to increase 
the locality of memory accesses and facilitate HW prefetching.

We conduct experiments on the PROSITE protein database for FAs of up to 702 FA states to
evaluate performance and effectiveness of our proposed optimizations.
Evaluations have been conducted on a 4 CPU (64 cores) AMD Opteron 6378 system 
and a 2 CPU (28 cores, 2 hyperthreads per core) Intel Xeon E5-2697 v3 system.
The observed speedups over the sequential baseline algorithm are up to 118541x 
on the AMD system and 2113968x on the Intel system.
\end{abstract}

\begin{IEEEkeywords}
SFA construction; fingerprints; hashing; parallelization; multicores
\end{IEEEkeywords}

\IEEEpeerreviewmaketitle

\section{Introduction}
\label{sec:introduction}

Multicore architectures have 
become main-stream. The
ubiquity of SIMD units, multiple cores and GPUs in today's desktops,
servers and even handheld devices necessitates the adaptation of
standard algorithms to expose parallelism and utilize
parallel execution units. Parallelization improves performance 
and makes algorithms scalable to larger problem sizes.

Searching a pattern from a large text has been a prevalent processing
step for various applications in computer science. Examples include
text editors, compiler front-ends, scripting languages, web
browsers, internet search engines, and security and DNA sequence
analysis. FAs derived from regular expressions  enable
such
uses, but the underlying, sequential FA algorithm
has linear complexity in the size of the input.
Significant research effort has alredy been
spent on parallelizing FA matching~\cite{Ladner:1980,Hillis:1986,Misra:2003,ScarpazzaVP07,Holub:2009,Jones2009,Luchaup2009,WangHL10,Luchaup2011,Ko2014,Mytkowicz2014}.
FA matching has been proven hard to be parallelized and inefficient
on parallel architectures, due to an existing dependency between state
transitions, i.e., for an FA to perform the next state transition, the
result state from the previous transition must be known.

To speed up FA matching on parallel architectures, SFAs have been proposed
in~\cite{Sinya2013}.  Given an FA~$A$ with $n$~states, the corresponding
SFA~$S(A)$ simulates $n$ parallel instances of FA~$A$. (This simulation is
similar to the simulation of a non-deterministic FA by a deterministic FA
through the subset construction algorithm~\cite{HopcroftU79}.) In particular,
an SFA state~$s$ is a vector of dimension~$n$ of FA states; the SFA start-state
is the vector~$\langle q_0,\ldots,q_{n-1}\rangle$, where each $q_i$ is a state
of the original FA. An SFA transition from SFA state~$s_1$ to SFA state~$s_2$
on input symbol~$\sigma$, denoted by $s_1\overset{\sigma}{\rightarrow} s_2$,
subsumes the transitions of the underlying FA from each of the FA states in
vector~$s_1$. Running the SFA on a sub-string of the input constitutes
$n$~parallel executions of the corresponding FA, each from a unique state
from the set of $A$'s states.  Given an SFA, it is then possible to split the
input into sub-strings, match each of the sub-strings in parallel with the SFA,
and combine the result vectors by reduction.

SFA construction requires considerable
time and space in terms of the number of FA states.
Therefore, the algorithm mostly becomes intractable for real-world
problem sizes. In particular, we found
that a large part of the sequence patterns from the
PROSITE protein sequence database~\cite{PROSITE,ScanProsite:2002}
exceeded the computational power of a contemporary 4-CPU (64 physical cores)
multicore server with \SI{128}{\giga\byte} of main memory.

The objective of this research is to improve the efficiency of the SFA
construction algorithm, to make it tractable for real-world problem sizes.
We identify the most time- and space-consuming
part of the SFA construction algorithm, which is the representation
and comparison of SFA states. We introduce fingerprints
to represent SFA state (vectors) by a single 64-bit quantity,
which reduces both the program's working set and
the time spent for the comparison of SFA states.
We introduce a hash-table that hashes SFA states on their
fingerprints. This hash-table reduces the set membership
test of SFA states to O(1).
For our parallelization of the SFA construction algorithm we designed,
implemented, and evaluated several approaches to find the most efficient
combination of coarse-grained, medium-grained and fine-grained parallelism. Our
parallel SFA construction algorithm is non-blocking and we state the termination
condition for this algorithm.

This paper makes the following contributions:

\begin{enumerate}
\item {Fingerprints and x86 ISA supporting operations:} Performing a membership
test on a set of SFA states is an expensive computation due to the increasing
number of states during SFA construction. To speed up set membership tests on
SFA states, we employ Rabin fingerprints~\cite{Broder93, Rabin81}.  We adapt
the Barrett reduction algorithm~\cite{Barrett87} and use a special x86
instruction~\cite{IntelCRC, AVX2} to attain fast fingerprinting.

\item{Hashing:}
Determining whether an SFA state has already been generated
requires a comparison to all previously generated states.
Although the fingerprints suggested in this paper reduce
the amount of comparisons, the linear search is still
costly.
To reduce the number of comparisons to O(1), we introduce
hashing that exploits fingerprints as key values.

\item {Several parallelization approaches:}
Within the construction algorithm, there are several sources
of parallelization.
As proposed and presented in former research~\cite{Choi2013},
which targeted a parallel version of the subset construction
algorithm, we
investigate all possible parallelism sources and determine which benefit
SFA construction on multicore platforms. 
We devise a static work distribution method and optimize all data-structures
to increase the locality of memory accesses.

\item
We evaluate our parallel SFA construction algorithm and SFA-based FA matching
for a selection of DNA sequence patterns from the PROSITE protein sequence
database~\cite{PROSITE,ScanProsite:2002}.  Our evaluation platforms include a
4-CPU (64 physical cores) AMD Opteron system and a 2-CPU (28 cores, 2
hyperthreads per core) Intel Haswell E5-2697 v3 system.
\end{enumerate}

The remainder of this paper is organized as follows.  In
Section~\ref{sec:background} we explain the relevant background material, specifically the
sequential SFA construction method and Rabin fingerprints.  In
Section~\ref{sec:optimizationSFA} we present our optimization methods for the
SFA construction algorithm.  Section~\ref{sec:experimentalResults} contains our
experimental evaluation. We discuss the related work in
Section~\ref{sec:relatedWork} and draw our conclusions in
Section~\ref{sec:conclusions}.

\section{Background}
\label{sec:background}
{\bf Finite automata:} A tuple $(\StateSets, \Sigma, \TF, \ensuremath{I}, F)$ describes a
deterministic finite automaton (DFA)~$A$. $\StateSets$ is a finite set of
states and $\NrStates$ is
referred to as the size of the DFA. $\Sigma$ is a finite alphabet of
characters and
$\Sigma^*$ is the set of strings over $\Sigma$.
$\ensuremath{I}\subseteq\StateSets$ is a set of initial states, but
in DFAs there is one initial state $\StartState\in\StateSets$, called
the start state.
$F \subseteq
\StateSets$ is the set of accepting states.
$\TF$ is a transition function of $\StateSets
\times \Sigma \to \StateSets$.
We extend transition function~$\TF$ to $\TFE$:
$\TFE(\State, ua)=p\Leftrightarrow\TFE(\State,u)=\State'$,
$\TF(\State',a)=p$, $a\in\Sigma$, $u\in\Sigma^*$.
An input string~$\STR$ over $\Sigma$ is accepted by DFA~$A$ if the DFA contains a labeled
path from $\StartState$ to a final state such that this path reads~$\STR$.
The DFA membership test is conducted by
computing~$\TFE(\State[0],\STR)$ and checking whether the result
is an accepting state.
As a notational convention, we denote the symbol
in the ~$i\,$th position of the input string by~$\STR[i]$.

\begin{figure}
\subcaptionbox{Example FA\label{fig:mot_ex:dfa}}[.5\textwidth]
{
    \begin{tikzpicture}[->,>=stealth']
        \tikzstyle{every initial by arrow}=[->]
        \node[smallstate, initial] (S0) {\state[0]};
        \node[smallstate, right of = S0, node distance = 4em] 
            (S1){\state[1]};
        \node[smallfinalstate, right of = S1, node distance = 3.4em]
            (S2){\state[2]};
        \path   (S0)    edge[loop above]    node{$\Sigma\hspace{0.4pt}\setminus\hspace{0pt}$$R$}   ()                   (S0)    edge[bend left]     node[above]{$R$}   (S1)
                (S1)    edge[bend left]     node[below]{$\Sigma\negthinspace\setminus\negthickspace\{R,G\}$}   (S0)
                (S1)    edge[loop above]    node{$R$}   ()
                (S1)    edge                node[above]{$G$}   (S2)
                (S2)    edge[loop above]    node{$\Sigma$}  ()
        ;
    \end{tikzpicture}
}
\subcaptionbox{transition table\label{fig:mot_ex:delta}}[.5\textwidth]
{
  \begin{tabular}{c|ccc}
    \hline
      $\delta$&$\Sigma\negthinspace\setminus\negthickspace\{R,G\}$&$R$&$G$\\
      \hline
      $0$&$0$&$1$&$0$\\
      $1$&$0$&$1$&$2$\\
      $2$&$2$&$2$&$2$\\
     \hline
  \end{tabular}
}
\subcaptionbox{sequential matching routine on input~\STR.\label{fig:mot_ex:match}}[.5\textwidth]
{
    \fbox{
        \begin{minipage}{.33\textwidth}
            \begin{internallinenumbers}
            state$\hspace{2pt}\leftarrow0$\\
            \textbf{for}$\hspace{2pt}i\leftarrow0\hspace{2pt}\textbf{to}\hspace{2pt}\lvert\STR\rvert-1\hspace{2pt}\textbf{do}$\\
            $\llcorner$\hspace{10pt}state$\hspace{2pt}\leftarrow\TF$(state,$\hspace{2pt}\STR[i]$)
            \end{internallinenumbers}
        \end{minipage}
    }
}
\caption{Example FA: state diagram, transition table and matching routine
}\label{fig:exDFA}
\end{figure}

Fig.~\ref{fig:mot_ex:dfa} shows an example FA over the
alphabet of one-letter abbreviations for the 20~amino-acids
($\Sigma=\{A,C,D,E,\allowbreak F,\allowbreak G,\allowbreak H,\allowbreak I,K,L,M,N,P,Q,R,\allowbreak S,\allowbreak T,\allowbreak V,\allowbreak W,Y\}$).
The FA accepts input strings that contain the sequence~$RG$ and has
start state~$\state[0]$ and a final state~$\state[2]$. The table in
Fig.~\ref{fig:mot_ex:delta} encodes the transition function~$\delta$,
and Fig.~\ref{fig:mot_ex:match} shows a sequential matching routine.

\begin{figure}[ht]
\subcaptionbox{Constructed SFA from FA in Fig.~\ref{fig:mot_ex:dfa}\label{fig:exSFA:dfa}}
{
    \begin{tabular}{ccc}
    \begin{tikzpicture}[->,>=stealth']
        \tikzstyle{every initial by arrow}=[->]
        \node[smallstate, initial] (S0) {$f_0$};
        \node[smallstate, right of = S0, node distance = 8em]
            (S2){$f_2$};
        \node[smallstate, above of = S2, node distance = 4em]
            (S1){$f_1$};
        \node[smallstate, below of = S2, node distance = 4.5em]
            (S3){$f_3$};
        \node[smallfinalstate, right of = S1, node distance = 7em]
            (S4){$f_4$};
        \node[smallstate, right of = S3, node distance = 7em]
            (S5){$f_5$};
        \path   (S0)    edge                node[above]{$R$}    (S1)
                (S0)    edge                node[above,xshift=0.8mm,yshift=-0.8mm]{$\Sigma\negthinspace\setminus\negthickspace\{R,G\}$}   (S2)
                (S0)    edge                node[above]{$G$}    (S3)
                (S1)    edge[loop above]    node{$R$}   ()
                (S1)    edge                node[above]{$G$}    (S4)
                (S1)    edge[bend left]     node[right]{$\Sigma\negthinspace\setminus\negthickspace\{R,G\}$}   (S2)
                (S2)    edge[bend left]     node[left]{$R$} (S1)
                (S2)    edge[loop right]    node[right]{$\Sigma\hspace{0.4pt}\setminus\hspace{0pt}$$R$} ()
                (S3)    edge[loop above]    node[yshift=-1mm]{$\Sigma\hspace{0.4pt}\setminus\hspace{0pt}$$R$}    ()
                (S3)    edge[bend left]     node[above]{$R$}    (S5)
                (S4)    edge[loop above]    node[above]{$\Sigma$}   ()
                (S5)    edge                node[right]{$G$}    (S4)
                (S5)    edge[bend left]     node[below]{$\Sigma\negthinspace\setminus\negthickspace\{R,G\}$} (S3)
                (S5)    edge[loop right]    node[right]{$R$}    ()
        ;
    \end{tikzpicture}&
    \end{tabular}
}

\subcaptionbox{State mapping table\label{fig:exSFA:delta}}[.5\textwidth]
{
    \begin{tabular}{|c|c||c|c|}
        \hline
        &$0\rightarrow\{0\}$&&$0\rightarrow\{0\}$\\
        $f_0$&$1\rightarrow\{1\}$&$f_3$&$1\rightarrow\{2\}$\\
        &$2\rightarrow\{2\}$&&$2\rightarrow\{2\}$\\
        \hline
        &$0\rightarrow\{1\}$&&$0\rightarrow\{2\}$\\
        $f_1$&$1\rightarrow\{1\}$&$f_4$&$1\rightarrow\{2\}$\\
        &$2\rightarrow\{2\}$&&$2\rightarrow\{2\}$\\
        \hline
        &$0\rightarrow\{0\}$&&$0\rightarrow\{1\}$\\
        $f_2$&$1\rightarrow\{0\}$&$f_5$&$1\rightarrow\{2\}$\\
        &$2\rightarrow\{2\}$&&$2\rightarrow\{2\}$\\
        \hline
    \end{tabular}
}
\caption{Example SFA: state diagram and state mapping table}\label{fig:exSFA}
\end{figure}



\begin{algorithm}[ht]
    \SetKwInOut{Require}{Require}
    \SetKwInOut{Ensure}{Ensure}
    \SetKwData{STATE}{state}
    \SetKw{RETURN}{return}
    \Require{Automaton $\mathcal{A}=(\StateSets,\Sigma,\TF,\ensuremath{I},\FinalStates)$}
    \Ensure{SFA $\mathcal{S}=(\StateSet[s],\Sigma,\TF_s,\IState[s],\FinalStates_s)$ is equivalent to an automaton $\mathcal{A}$}
    $\StateSet[s]\leftarrow\emptyset, \StateSet[tmp]\leftarrow\{f_I\}$\\
    \While{$\StateSet[tmp]\neq\emptyset$}{
        choose and remove a mapping $f$ from $\StateSet[tmp]$\\
        $\StateSet[s]\leftarrow\StateSet[s]\cup\{f\}$\\
        \ForAll{$\sigma\in\Sigma$}{
            $\ensuremath{q}\in\StateSets\hspace{7pt}f_{next}(\ensuremath{q})\mathrel{\mathop:}=\bigcup_{q'\in f(q)}\TF(q',\sigma)$\\
            $\TF_s[f,\sigma]\leftarrow f_{next}$\\
            \If{$f_{next}\notin\StateSet[s],\StateSet[tmp]$}{
                $\StateSet[tmp]\leftarrow\StateSet[tmp]\cup\{f_{next}\}$
            }
        }
    }
    $\IState[s]\leftarrow\{f_I\}$\\
    $\FinalStates_s\leftarrow\{f\in\StateSet[s]|\exists q\in\IStates|f(q)\cap\FinalStates\neq\emptyset\}$
    \caption{SFA Construction}
    \label{algo:constructSFA}
\end{algorithm}

{\bf Simultaneous Deterministic Finite Automata:}
Due to the fundamental reason that the overhead of speculation is
inevitable for parallel FA matching, SFA is introduced
\cite{Sinya2013}. SFA is a model of efficient parallelization for FA
matching, generated by the original FA. Algorithm~\ref{algo:constructSFA} denotes the
construction algorithm to develop SFAs. Details of the algorithm can be found in \cite{Sinya2013}.

Fig.~\ref{fig:exSFA:dfa} is an example SFA and
Fig.~\ref{fig:exSFA:delta} is the corresponding state mapping table,
generated by the FA in Fig.~\ref{fig:mot_ex:dfa}.
In summary, the construction algorithm keeps finding new SFA-states, checking
whether they are already in a set and adding them if they are not in
the set yet, starting with the set that has only the start state $f_I$.
After the algorithm is finished, a SFA and a corresponding state mapping
table are finally
produced. For the example SFA, $\StateSet[s]$, whose elements in
Fig.~\ref{fig:exSFA:dfa} are $f_i$, $0\le i \le5$.

{\bf Rabin Fingerprints:}
Fingerprints are short bit-strings for larger objectes. If two
fingerprints are different, the corresponding objects are known
to be different. There is a small probablility that two
objects map to the same fingerprint, which is called a collision.
Michael O.~Rabin's fingerprinting method~\cite{Rabin81,Broder93}
creates fingerprints from arbitrary
bit-strings
$A=\{a_1,a_2,\cdots,a_m\}$ by interpreting $A$ as
a polynomial $A(t)$ of degree $m-1$ with coefficients
in $\mathbb{Z}_2$. 

\begin{equation}
\centering
\begin{tabular}{c}
    $\mathcal{A}(t)=a_1t^{m-1}+a_2t^{m-2}+\cdots+a_m.$
\end{tabular}
\label{eq:inputA}
\end{equation}
To create a $k$-bit fingerprint, an irreducible random polynomial
$\mathcal{P}(t)$ over $\mathbb{Z}_2$ of degree~$k$ is selected.
The Rabin fingerprint can then be defined as 
\begin{equation}
\centering
\begin{tabular}{c}
    $f(A)=\mathcal{A}(t)$~mod~$\mathcal{P}(t).$ \\
\end{tabular}
\label{eq:rabin}
\end{equation}

\noindent As shown in~\cite{Broder93}, the probability of a collision among a
set of $n$ distinct bit-strings is less than $\frac{n^2m}{2^k}$.

To apply Rabin fingerprints, the costs for computing the
polynomial modulo operation from Eq.~\eqref{eq:rabin} must
be kept to a minimum. 
The Barret reduction method from~\cite{Barrett87} can be used
to express computationally costly modulo operations
in terms of multiplication, division, addition and the
floor function.
For arbitrary integers $a$ and $n$, the basic idea of Barrett reduction is
denoted as
\begin{equation}
\centering
\begin{tabular}{c}
    $a$ mod $n=a-\lfloor am\rfloor n$, where $m=\sfrac{1}{n}$. \\
\end{tabular}
\label{eq:barret}
\end{equation}

Rabin's fingerprinting method $\mathcal{A}(t)$
mod $\mathcal{P}(t)$ can be calculated through Barrett reduction as

\begin{equation}
\centering
\begin{tabular}{ll}
    $\mathcal{A}(t)$ mod $\mathcal{P}(t)=$&$\mathcal{A}(t)\oplus
    \Big\lfloor (\lfloor\sfrac{\mathcal{A}(t)}{t^k}\rfloor\bullet
    \lfloor\sfrac{t^{2k}}{\mathcal{P}(t)}\rfloor)\div{t^k}\Big\rfloor$\\
    &$\bullet\mathcal{P}(t)$,
\end{tabular}
\label{eq:rabinBarrett}
\end{equation}
where
$\oplus$ and $\bullet$ denote 
addition and multiplication in the Galois Fields of
characteristic 2 (GF($2^k$)).
In particular, $\bullet$ is
a carry-less multiplication operation.

\begin{equation}
\centering
\begin{tabular}{l}
    T1pre~$=\lfloor\sfrac{\mathcal{A}(t)}{t^k}\rfloor$,
    ~M~$=\lfloor\sfrac{t^{2k}}{\mathcal{P}(t)}\rfloor$\\
    T1~$=$~T1pre$~\bullet~$M\\
    T2pre~$=\lfloor$~T1$~\div~t^k\rfloor$\\
    T2~$=~$T2pre$~\bullet~\mathcal{P}(t)$\\
    $\mathcal{A}(t)$~mod~$\mathcal{P}(t)=\mathcal{A}(t)\oplus~$T2
\end{tabular}
\label{eq:rabinStep}
\end{equation}

Eq.~\eqref{eq:rabinStep} displays the work-flow of the fingerprinting
algorithm from Eq.~\eqref{eq:rabinBarrett} step by step.

The x86 architecture provides the {\tt PCLMULQDQ} SSE instruction to perform
carry-less multiplication efficiently in hardware (according to
\cite{IntelCarryless}, the savings of the {\tt PCLMULQDQ}
instruction are
100~cycles per multiplication).

The {\tt PCLMULQDQ} instruction takes two
\SI{64}{\bit} operands and returns a \SI{128}{\bit} result.
The SFA states we will fingerprint are mostly larger than \SI{64}{\bit}.
Even if we store
an FA state as type {\tt unsigned short} (\SI{16}{\bit}), only 4~states fit
within \SI{64}{\bit}. We apply the folding method
described in \cite{IntelCRC} to reduce SFA states of arbitrary
size to \SI{64}{\bit} fingerprints.

\section{Optimizing the SFA Construction Algorithm}
\label{sec:optimizationSFA}

\begin{equation}
\centering
\begin{tabular}{ll}
    $\mathcal{O}(\sum\limits_{i=1}^{|\StateSet[s]|}\sum\limits_{j=1}^{|\Sigma|}(|\StateSet|+|\StateSet|\times~i))=$&$\mathcal{O}(\frac{1}{2}\times|\Sigma|\times|\StateSet|\times|\StateSet[s]|$\\
    &$\times(|\StateSet[s]|+3))$.
\end{tabular}
\label{eq:timeComp}
\end{equation}

The time complexity of Algorithm~\ref{algo:constructSFA} can be determined
as follows; because the outmost while loop (line 2) keeps working until
no more elements are left in~$\StateSet[tmp]$,
while loop
iterates $|\StateSet[s]|$ time, one per SFA state, same as the size
of the resulting SFA. Inside of while loop, next states for each
SFA state are calculated on each symbol one by one in for loop (line 5). Thus
for loop iterates $|\Sigma|$ times per SFA state. To decide
whether a newly created state is already in the set $\StateSet[s]$,
the ``exhaustive" set membership test (comparing all FA states in SFA state)
between the new state
and all other states in sets
should be done. If we presume in the worst-case that all states are different,
set membership test should be done as same as the number of
states in the set at that moment and each membership test need
$|\StateSet|$ comparisons of FA state. Therefore, the time complexity of
this algorithm on worst-case is given in the equation~\eqref{eq:timeComp}.

From the time complexity, we can recognize that the number of generated
SFA states is the most significant factor for time consumption
of this algorithm. As we can see,
comparisons for newly generated states take the
most of
the time $\mathcal{O}(|\Sigma|\times |\StateSet|\times|\StateSet[s]|\times
(|\StateSet[s]|+1)\times \frac{1}{2})$. The part for checking termination
$\mathcal{O}(|\StateSet[s]|)$ and state transition $\mathcal{O}(|\StateSet[s]|
\times|\Sigma|\times|\StateSet|)$ only take a very small portion of the whole algorithm. This
means we need to focus on decreasing time taken by the state comparison
part in the following optimization.

\subsection{Hashing with Fingerprints}
If we adapt fingerprints, then all states already have fingerprints and they will just compare the value of fingerprints with each other. The exhaustive matching is only conducted when two fingerprints from states are identical.
However, without fingerprints, states will compare their own contents with each other by exhaustive comparison.
As a result, our algorithm is the non-probabilistic approach which always constructs 
the exact SFA.

By exploiting fingerprints further, we introduce a hash table data
structure to our construction algorithm. This can be done easily
because fingerprints are stored in an~{\tt unsigned int}~or~{\tt unsigned long long}~data type
({\tt uint32\_t} or {\tt uint64\_t}) so that we can use it directly
as a key value for hashing.

We declare the hash table as
a pointer array to indicate each SFA state node. During
construction algorithm, a fingerprint for each generated state is
calculated before the set membership test. Each hash table entry uses this
information to point out a SFA state node which has corresponding
fingerprint value.

If there is no duplication in fingerprints, the generated state is
added to the hash table at the entry of hash index. Otherwise,
the new state should follow the pointer chain from its hash entry.
While going to the leaf of linked list pointer chain, comparison with
states in the chain needed to be done. If the new state cannot meet
the same state until it reaches the leaf node, it is added to the leaf
of the linked list chain. At the best case, now we only need $\mathcal{O}(1)$ time
to find states that have the same fingerprint.

\subsection{Parallelization for Multicore Architectures}
\label{sec:parallelOnMulti}

\subsubsection{Sources of Parallelism}
\label{subsubsec:parSources}
As presented in the work of parallelizing the subset
construction algorithm~\cite{Choi2013}, the sequential SFA construction
algorithm presented in Algorithm~\ref{algo:constructSFA} also has
similar parallel sources in itself, from coarse-grained parallelism to
fine-grained parallelism.

The fine-grained parallel source is not practical. We could
divide state transitions (line 6) of FA states but total work amount
for state transition is rather small ($|\StateSet[s]|\times|\StateSets|$)
compared to the state comparison part to gain some recognizable benefit. Also, it's questionable
how we can parallelize it smart enough when the number of participating
processor exceeds the number of FA states. It is hard to distribute and
assign work to processors without making idle processors. Moreover,
after splitting the work, the partial result of a transition should be gathered
to make a new SFA state. It means introduction of synchronization and waiting overhead between
threads. We might gain some advantage from the cache locality
because transitions are split based on the FA states and one row in the transition
table $\TF$ corresponds to the transition information of one FA state.
But still, compared what we could gain from the fine-grained parallelization
to the overhead of synchronization, this approach is hard to be a good option
for task-level parallelization.

The most significant and easy to recognize source is coarse-grained
parallelism which is dividing the work based on the SFA states. In
this case, the while loop (from line 2 to 9) is parallelized. However,
this parallelization can be meaningful only if unprocessed states in
the $\StateSet[tmp]$ are always in a plenty number so that all processors
can have a state to process in most of time during construction and not to
go into idle state. This approach would be appropriate for FAs which have a large number of states.

Medium-grained parallelism is dividing tasks based on symbols.
Each thread can do series of works by itself, without help of other threads;
producing new states on given symbols, testing the possible duplication of states 
and adding them to the set. 
Now synchronization is not needed because each thread can proceed individually.
However, still we need a good scheme to distribute work to processors
so that every processor has the same amount of work. Also, consideration of when 
the number of processors is greater than the number of symbols should be needed.

We utilize coarse-grained parallelism for the transposed transition table and medium-grained parallelism for static work allocation.

\subsubsection{Static Work Allocation}
\label{subsubsec:Static}
\begin{algorithm}[htp]
    \SetKwInOut{Require}{Require}
    \SetKwInOut{Ensure}{Ensure}
    \SetKwData{STATE}{state}
    \SetKw{RETURN}{return}
    \Require{Automaton $\mathcal{A}=(\StateSets,\Sigma,\TF,\ensuremath{I},\FinalStates),~|P|=$the number of threads,\\Distribution of$~\Sigma,~D=\{B_0,\cdots,B_{P-1}\}$,\\current threads: $j$th thread}
    \Ensure{SFA $\mathcal{S}=(\StateSet[s],\Sigma,\TF_s,\IState[s],\FinalStates_s)$ is equivalent to an automaton $\mathcal{A}$}
    $\StateSet[s]\leftarrow\{f_I\}, Index_j\leftarrow0$\\
    \WhilePar{$\exists\StateSet[s][Index_j]\neq\emptyset$}{
        \While{$\forall\sigma_i\in~B_j$}{
            choose a mapping $f$ from $\StateSet[s][Index_j]$\\
                    $\ensuremath{q}\in\StateSets\hspace{7pt}f_{next}(\ensuremath{q})\mathrel{\mathop:}=\bigcup_{q'\in f(q)}\TF(q',\sigma_{i})$\\
                    $\TF_s[f,\sigma_{i}]\leftarrow f_{next}$\\
                    \If{$f_{next}\notin\StateSet[s]$}{
                        $\StateSet[s]\leftarrow\StateSet[s]\cup\{f_{next}\}$\\
                    }
                    $Index_j\leftarrow Index_j+1$\\
        }
    }
    $\IState[s]\leftarrow\{f_I\}$\\
    $\FinalStates_s\leftarrow\{f\in\State[s]|\exists q\in\IStates|f(q)\cap\FinalStates\neq\emptyset\}$
    \caption{Parallel SFA Construction with static distribution (fewer threads than symbols)}
    \label{algo:parConstStaticSmall}
\end{algorithm}

Algorithm~\ref{algo:parConstStaticSmall} describes the parallel algorithm
with the static distribution of symbols when the number of threads $|P|$ is
less than the number of symbols $|\Sigma|$. An array that has distribution information for each thread is introduced.
Because the number of threads is less than the number of symbols, every thread
has to compute one or more symbols and we need a variable
to hold those data. A vector $D=\{B_0,~\cdots,~B_{P-1}\}$
has the distribution information of symbols $B_j$ about what symbols are assigned
to the thread $j$. Thus, iteration for symbols (line 3) is limited to only
assigned symbols in $B_j$. The inside of iteration, lines from 4 to 9, each thread computes next nodes with assigned symbols. When newly created
state $f_{next}$ is not in the set $\StateSet[s]$, it is added to $\StateSet[s]$.
This algorithm finishes when all threads do not have any non-processed SFA state (line 2).

\begin{algorithm}[ht]
    \SetKwInOut{Require}{Require}
    \SetKwInOut{Ensure}{Ensure}
    \SetKwData{STATE}{state}
    \SetKw{RETURN}{return}
    \Require{Automaton $\mathcal{A}=(\StateSets,\Sigma,\TF,\ensuremath{I},\FinalStates),~g=~$the number of groups,\\Partition of$~\StateSet[s],~D=\{S_0,\cdots,S_{g-1}\},$\\threads: $j$th thread in $k$th group}
    \Ensure{SFA $\mathcal{S}=(\StateSet[s],\Sigma,\TF_s,\IState[s],\FinalStates_s)$ is equivalent to an automaton $\mathcal{A}$}
    $\StateSet[s]\leftarrow\{f_I\}, Index_{k*g+j}\leftarrow0$\\
    \WhilePar{$\exists\StateSet[s][Index_{k*g+j}]\neq\emptyset$}{
        \While{$~\StateSet[s][Index_{k*g+j}]\in~S_k$}{
            choose a mapping $f$ from $\StateSet[s][Index_{k*g+j}]$\\
                $\ensuremath{q}\in\StateSets\hspace{7pt}f_{next}(\ensuremath{q})\mathrel{\mathop:}=\bigcup_{q'\in f(q)}\TF(q',\sigma_{j})$\\
                $\TF_s[f,\sigma_{j}]\leftarrow f_{next}$\\
                \If{$f_{next}\notin\StateSet[s]$}{
                    $\StateSet[s]\leftarrow\StateSet[s]\cup\{f_{next}\}$\\
                }
                $Index_{k*g+j}\leftarrow Index_{k*g+j}+1$\\
        }
    }
    $\IState[s]\leftarrow\{f_I\}$\\
    $\FinalStates_s\leftarrow\{f\in\State[s]|\exists q\in\IStates|f(q)\cap\FinalStates\neq\emptyset\}$
    \caption{Parallel SFA Construction with static distribution by grouping}
    \label{algo:parConstStaticGroup}
\end{algorithm}

We also introduce an algorithm for the opposite case
when the number of threads is greater than the number of symbols. In 
this case, we can consider two cases, whether the
number of threads is a multiple of the number of symbols or not. The
following Algorithm~\ref{algo:parConstStaticGroup} explains the
former case. In this case, because the number of threads is a multiple
of the number of symbols, $|\Sigma|$ threads are gathered together
to form a group. By doing this, each 
thread in a group will only take one symbol to process. Thus, SFA
states set $\StateSet[tmp]$ should be split into $S_0,~\cdots,~S_{g-1}$ and assigned to each group,
where $g$ is the number of groups. 

Now an array D holds distribution information of SFA states for each group
from $S_0~to~S_{g-1}$ when the number of groups is $g$. Threads in the group
$k$ will only choose a SFA state in the distribution set $S_k$. Also because
the number of threads in a group is the same as the number of symbols $|\Sigma|$,
the $j$th thread in a group will only compute next states on symbol $\sigma_j$.
Please note that there is no for loop that iterates over symbols, unlike in the 
previous algorithm. In other words, each group just cares about SFA
states assigned to them and each thread also computes the only symbol given to
them. Thus, a thread in the $k$th group starts its execution if there is a state
in $S_k$ that is not processed for the symbol which is given to it (line 3).
The algorithm between line 4 to 9 is the same as in the previous algorithm.

For the last condition, when the number of threads is not a multiple of but
greater than the number of symbols, the parallel algorithm with static work
distribution is a mixture of the above two Algorithms~\ref{algo:parConstStaticSmall}
and~\ref{algo:parConstStaticGroup}. In this case, the number of groups $g$ will be
$\lceil\frac{|P|}{|\Sigma|}\rceil$ where $|P|$ is the number of threads
and $|\Sigma|$ is the number of symbols. Among them, groups of
full threads which have $|\Sigma|$ threads will act as
presented in Algorithm~\ref{algo:parConstStaticGroup}. (The number of
such groups will be
$\big\lceil\frac{|p|}{|\Sigma|}\big\rceil-\big\lfloor\frac{|p|}{|\Sigma|}\big\rfloor$.)
The remaining groups which do not have enough threads to
have one symbol per thread, will follow Algorithm~\ref{algo:parConstStaticSmall}
except for line 2, because they also choose a state $f$ in the set $S_{g-1}$,
which is the partition of $\StateSet[tmp]$ for the last group.
Thus, threads in the last group follow Algorithm~\ref{algo:parConstStaticGroup} only for
line 2. Therefore, especially in this case, the distribution of SFA states
in $\StateSet[s]$ should be done by considering the number of threads for each group.

\subsubsection{Transposing Transition Tables}
\label{subsubsec:Transposition}
Transition tables of FAs are represented as 2-dimensional arrays
in row-major layout.  With those
transition tables, each thread has a high probability to access different rows
of them whenever it computes the next FA state according to the current FA
state and symbol. This decreases the locality of memory accesses and
eventually aggravates performance of SFA construction.

\begin{figure}[htp]
\centering
    source SFA state $s_0$=\{$\StateSet[1], \StateSet[0], \StateSet[2]$\}
    \includegraphics[height=30mm,draft=false,clip=false]{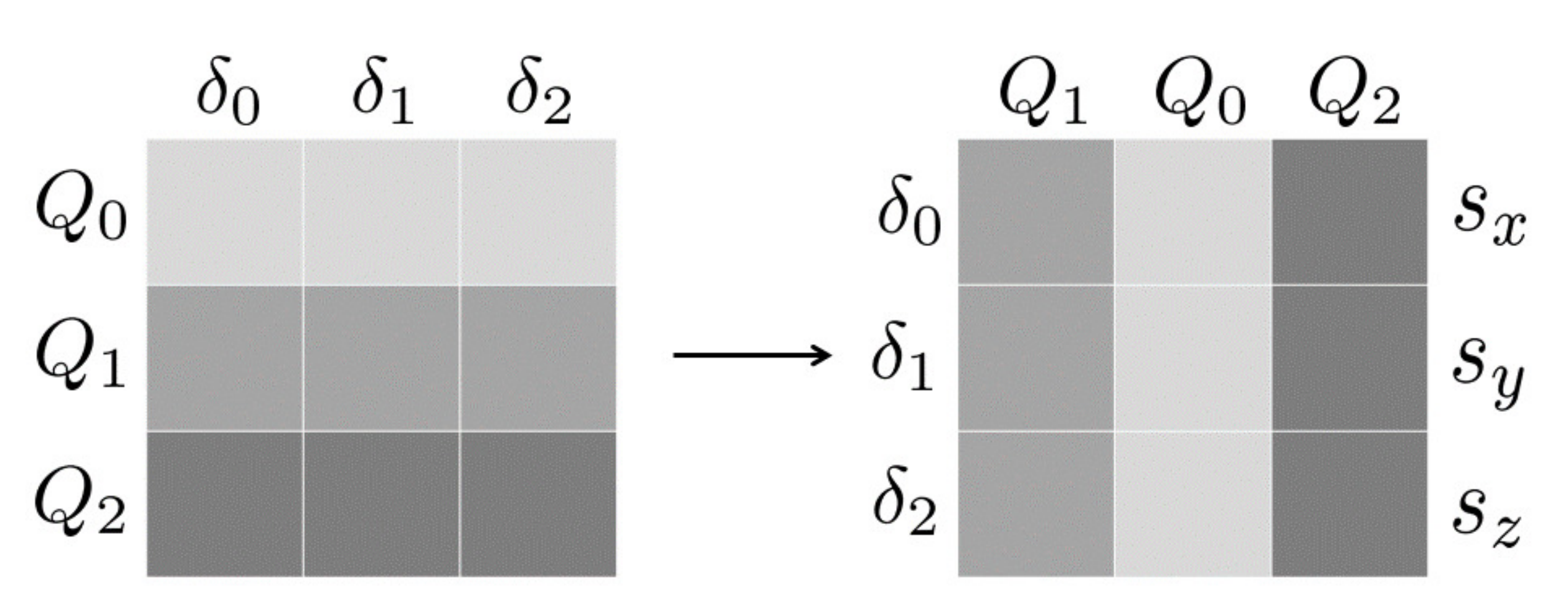}
\caption{Example transposed transition table}
\label{fig:transposedTable}
\end{figure}

As an example, suppose our algorithm is going to derive new SFA states
$\{s_x, s_y, s_z\}$ from a given SFA state~$s_0$ on a set of input
symbols~$\{\delta_0, \delta_1, \delta_2\}$. We transpose the transition
table~$\delta$ according to the DFA states in~$s_0$.  As follows from the left
part of Fig.~\ref{fig:transposedTable}, the transition table is read in
row-major order, with each row corresponding to one SFA state in~$s_0$. Each
such row becomes one column in the transposed table in the right part of
Fig.~\ref{fig:transposedTable}.  Each {\em row\/} in the transposed table
represents a new SFA state (which will again be processed in row-major order,
to facilitate locality).

The sequential algorithm of this approach is the same as Algorithm~\ref{algo:constructSFA} except 
for line 6 is replaced by transposing. Because transposing a transition table produces next SFA states
according to all symbols, coarse-grained parallelism is appropriate for this approach. 
Therefore, each thread takes one SFA state from the list of non-computed states until every thread
has no SFA states to process.

\section{Experimental Results}
\label{sec:experimentalResults}
In this section, we demonstrate how all optimization methods presented
in Section~\ref{sec:optimizationSFA} affect the 
performance of SFA construction algorithm.
\begin{table}
    \centering
    \begin{tabular}{|m{0.9cm}|m{2cm}|c|c|c|}
    \hline
    Name & CPU Model & CPUs & $\frac{\mbox{Cores(HTs)}}{\mbox{CPU}}$ & Clock Freq.\\
    \hline
    \hline
    Intel Xeon & Intel\hspace{0.1cm}Core\hspace{1cm}E5-2697 v3 & 2 & 14(28) & 2.60~GHz\\
    \hline
    AMD Opteron  & AMD\hspace{0.1cm}Opteron Processor 6378 & 4 & 16 & 2.40~GHz\\
    \hline
    \end{tabular}
    \caption{Hardware Specifications}
    \label{tab:HWSpec}
\end{table}
\noindent
We implemented SFA construction and matching for two architectures summarized
in Table~\ref{tab:HWSpec}. POSIX threads~\cite{Butenhof97} were used to
parallelize SFA construction and matching across multiple cores.  To generate
minimal DFAs from regular expressions, we use Grail+~\cite{Grail95,Grail}.
Our SFA construction and matching frameworks read DFAs and input strings in
Grail+ format and convert them to our framework's internal representation.

From the smallest benchmark with 5 DFA states to the largest benchmark with 2930
DFA states, total 1062 DFAs from PROSITE protein database~\cite{PROSITE} are used.

The interpolated lines of diagrams were created using R's local regression method.

\subsection{Optimizations for Sequential Algorithm}
\label{sec:sequential}
\begin{figure}[htp]
    \centering
    \vspace{-1mm}
    \subcaptionbox{on the AMD system\label{fig:seqElc2}}
    {
        \includegraphics[height=40mm,draft=false,clip=false]{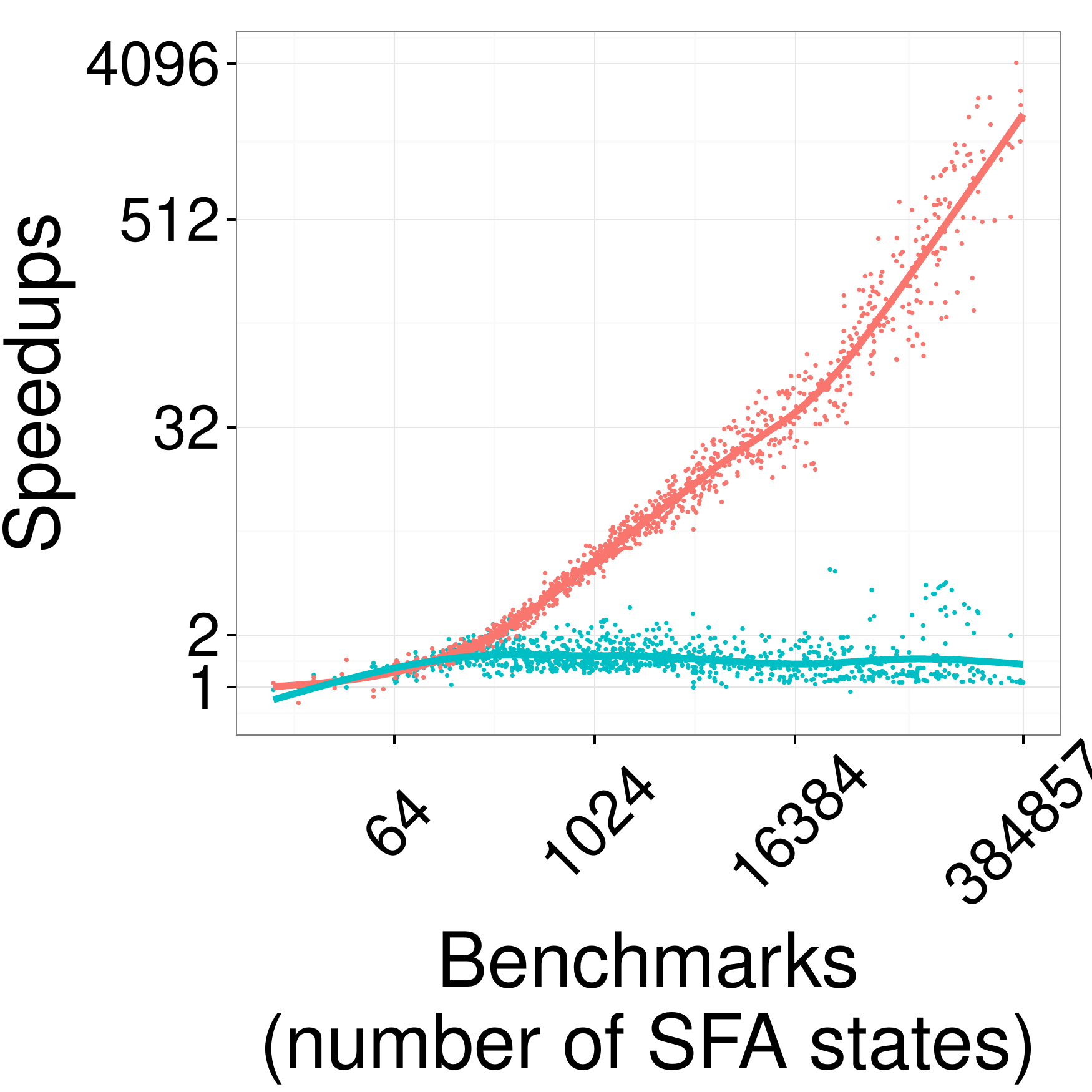}
    }\hfill
    \subcaptionbox{on the Intel system\label{fig:seqElc3}}
    {
        \includegraphics[height=40mm,draft=false,clip=false]{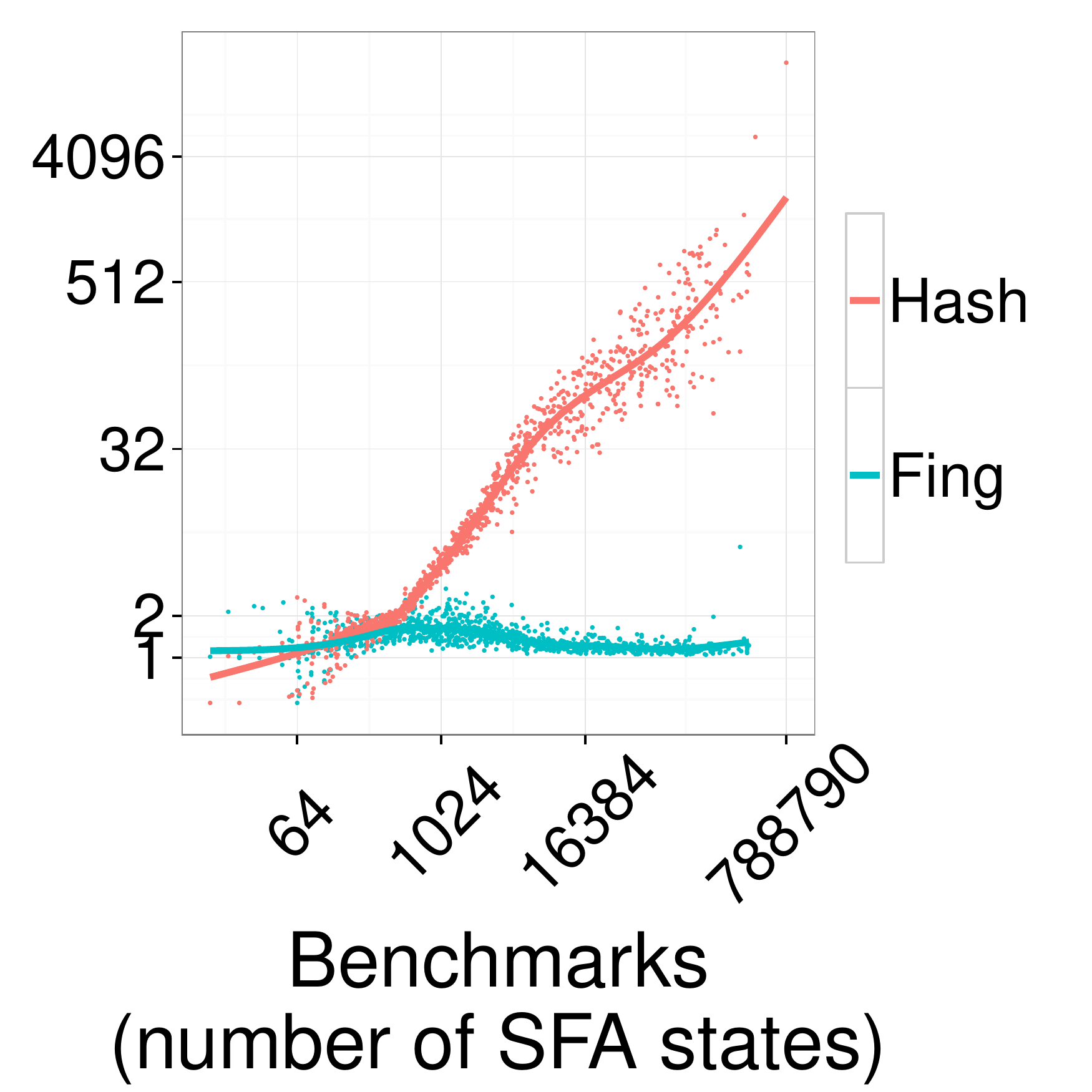}
    }
    \caption{Performance of fingerprints and hashing}
    \label{fig:seqExp}
    \vskip-0mm
\end{figure}
Figure~\ref{fig:seqExp} illustrates the speedup of fingerprints compared to the non-optimized 
baseline and the speedup of hashing compared to fingerprints. Therefore, the total speedup is
the multiplication of both of them. 

Because of time limit, it is impossible to experiment all benchmarks for the squential
methods, especially for the baseline.
Without fingerprints and hashing, the execution time for one benchmark might take over one day.
And we show the speedups of benchmarks which are tractable on our servers.
This explains the reason why we need to parallelize the sequential SFA construction.

\subsection{Parallelization}
\label{sec:parallel}
\begin{figure}[htp]
    \centering
    \vspace{-1mm}
    \subcaptionbox{on the AMD system\label{fig:parElc2}}
    {
        \includegraphics[height=70mm,draft=false,clip=false]{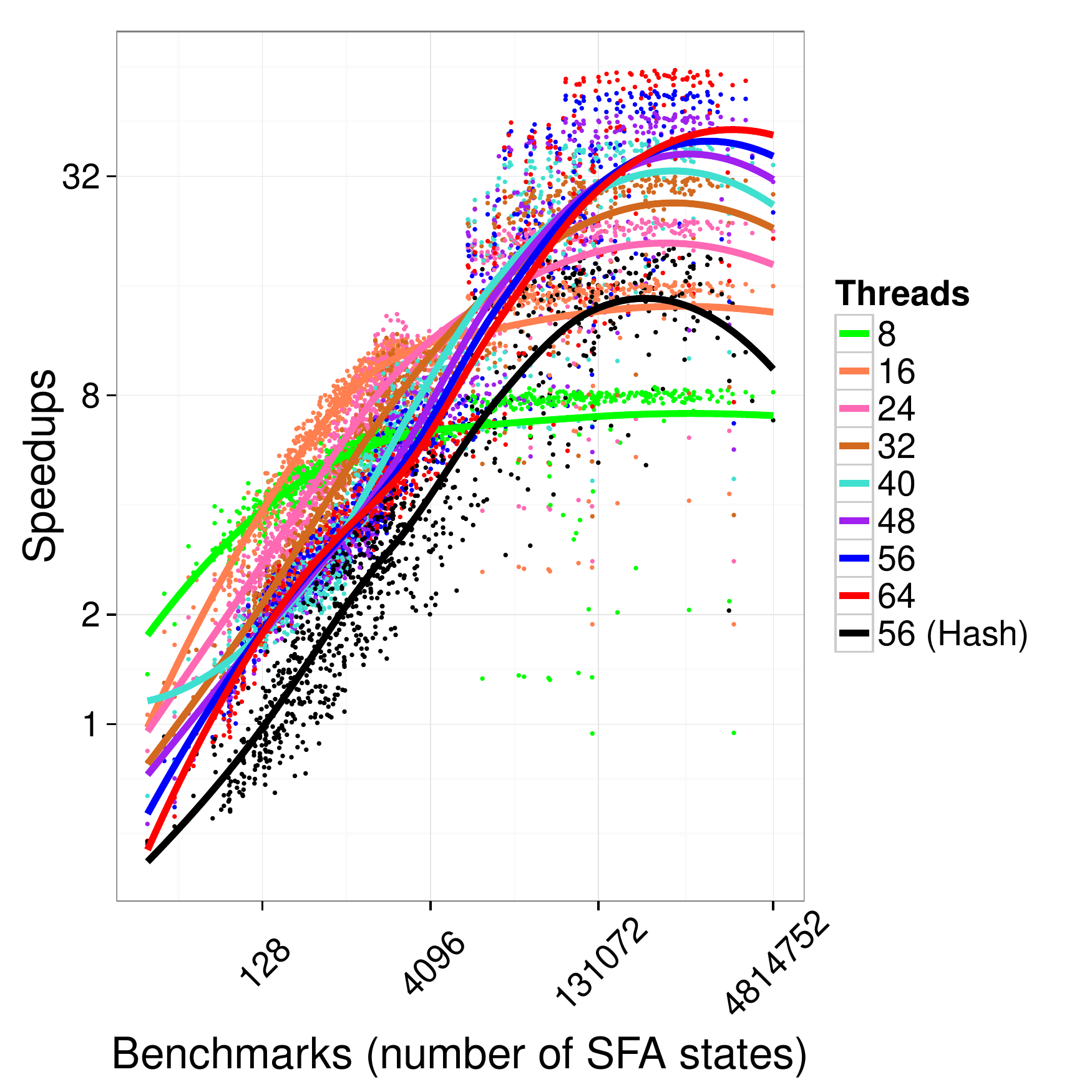}
    }\hfill
    \subcaptionbox{on the Intel system\label{fig:parElc3}}
    {
        \includegraphics[height=70mm,draft=false,clip=false]{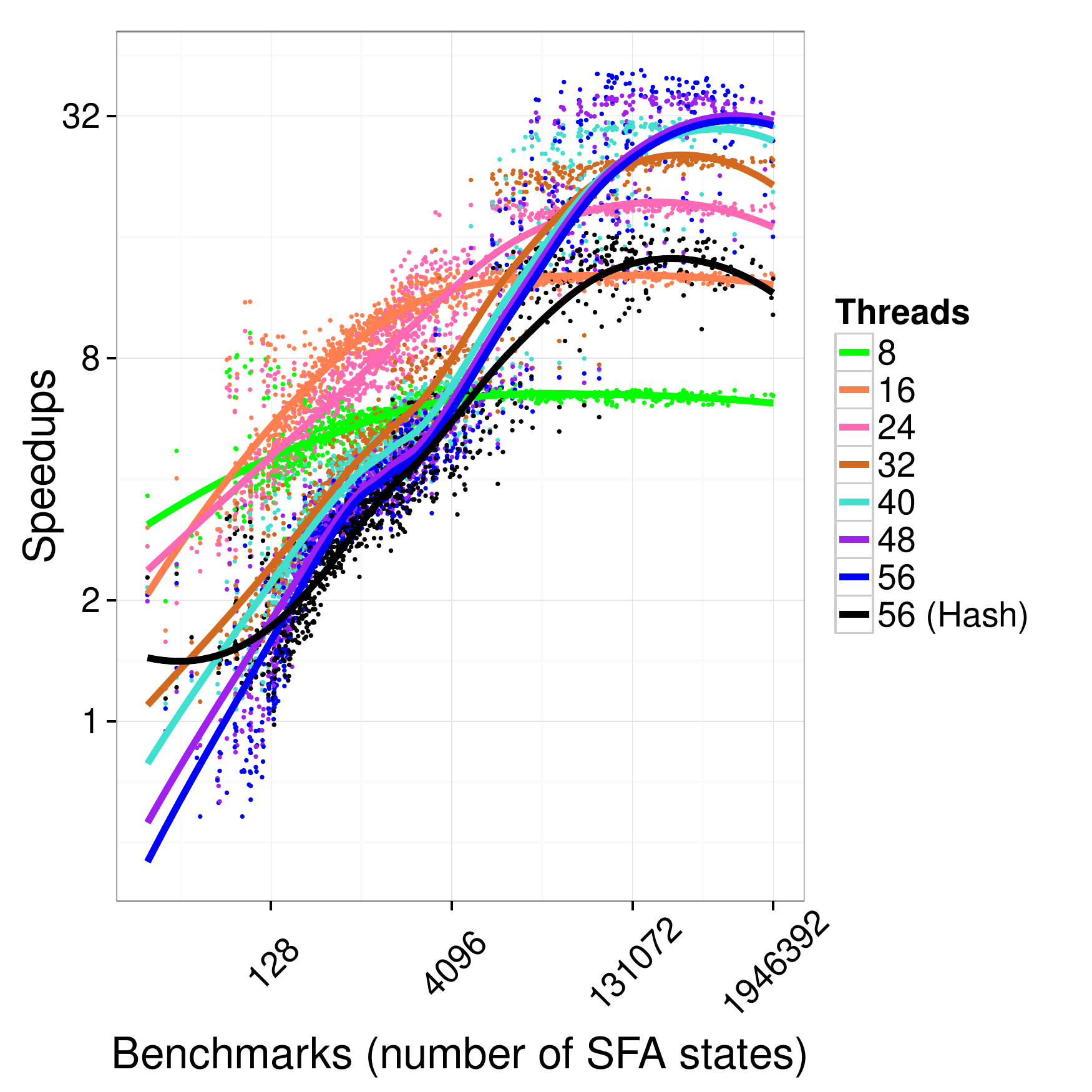}
    }
    \caption{Speedup of the best parallel implementation
             over the best sequential implementation.}
    \label{fig:parExp}
    \vskip-0mm
\end{figure}
Figure~\ref{fig:parExp} describes the speedup of our parallel SFA construction algorithm 
with fingerprints, hashing and transposing over the best optimized sequential implementation. 
Thus, the overall speedup compared to the baseline sequential algorithm can be achieved by
multiplying the parallel speedup, the sequential hashing speedup and the sequential fingerprints
speedup. 

Because the Intel system has only half (\SI{64}{\giga\byte})
of the memory size of the AMD
system, the range of benchmarks we could run are different depending on the
system.  The maximum speedups achieved by parallelization on the AMD and the
Intel
system are 62.365x with 64 threads and 41.438x with 56 threads while the
median speedups of them are 3.968x and 3.911x respectively. The reason why
median speedups are different from the maximum speedups is that there is not
enough work to distribute in the case of small benchmarks which generate
smaller numbers of SFA states. In that situation, a smaller number of threads
tends to show better performance, because more threads incur a penalty from
inter-core cache coherence transfers (larger SFA sizes do not fit into
the CPU caches anymore and rely more on data from main memory).

\subsection{SFA Matching}
\label{sec:sfaMatching}
\begin{figure}[htp]
    \centering
    \vspace{-1mm}
    \includegraphics[height=40mm,draft=false,clip=false]{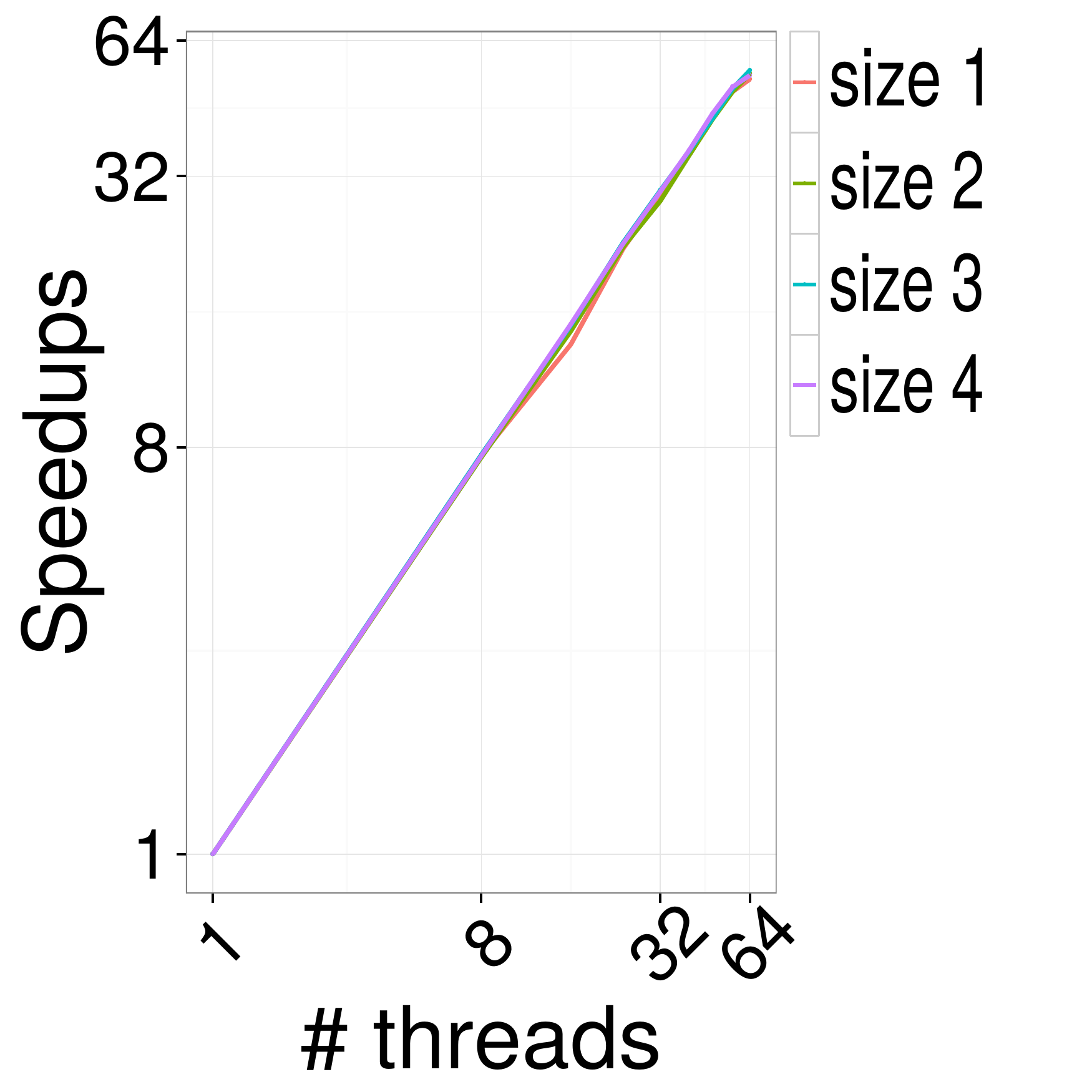}
    \caption{SFA matching}
    \label{fig:SFAMatch}
    \vskip-0mm
\end{figure}

The result of SFA matching with an input string of ten billion characters is
shown in  Figure~\ref{fig:SFAMatch}. The experiment has been conducted on the
64-core AMD Opteron system and we picked the median execution time of three
iterations for each benchmark.  The numbers of SFA states are 12299, 25332,
206351 and 2437146 respectively (size 1, size 2, size 3 and size 4 in
Figure~\ref{fig:SFAMatch}).  Because there is no dependency between the input
chunks, the speedup of SFA matching over sequential matching is linear in the
number of threads.
It is worth noting that this speedup has been observed even
with the large, 2,437,146-state SFA. The transition table of
this SFA has a size of \SI{97}{\mega\byte}, which exceeds
the \SI{12}{\mega\byte} L3 cache of each of the 4 Opteron 6378
CPUs already considerably. We thus expect that SFA matching will
scale to even larger SFAs.

\section{Related work}
\label{sec:relatedWork}
To the best of our knowledge, there are no challenges for parallelizing
the SFA construction so far. In this section, we only can 
introduce research efforts that do parallelization of FA matching,
which is based on speculation. 

Locating a string in a larger text has applications
with text editing, compiler front-ends and web browsers,
internet search engines, computer security, and DNA sequence analysis.
Early string
searching algorithms such as Aho--Corasick~\cite{Aho:1975},
Boyer--Moore~\cite{Boyer:1977} and Rabin--Karp~\cite{Karp:1987}
efficiently match a finite set of input strings
against an input text.

Regular expressions allow the specification of
infinite sets of input strings.
Converting a regular expression to a DFA for DFA membership tests is a
standard technique to perform regular expression matching.
The specification of virus signatures in intrusion
prevention systems~\cite{Brumley2006,Sommer2003,Roesch1999} and
the specification of DNA
sequences~\cite{sigrist2010prosite,boeckmann2003swiss}
constitute recent applications of regular expression matching with DFAs.

Considerable research effort has been
spent on parallel algorithms for DFA membership tests.
Ladner et al.~\cite{Ladner:1980} applied the parallel prefix computation
for DFA membership tests with Mealy machines.
Hillis and Steele~\cite{Hillis:1986} applied parallel prefix computations
for DFA membership tests on the 65,536 processor Connection Machine.
Ravikumar's survey~\cite{ParallelFA1998} shows how DFA membership tests
can be stated as a chained product of matrices.
Because of the underlying parallel prefix computation, all three approaches
perform a DFA membership test on input size~$n$
in $\mathcal{O}(\log(n))$ steps, requiring $n$~processors. Their algorithms
handle arbitrary
regular expressions, but the underlying assumption of a massive number
of available
processors can hardly be met in most practical settings.
Misra~\cite{Misra:2003} derived another $\mathcal{O}(\log(n))$ string
matching algorithm. The number of required processors is
on the order of the product of the two string lengths and hence not
practical.

A straight-forward way to exploit parallelism with DFA membership tests is
to run a single DFA on multiple input streams in parallel, or to run
multiple DFAs in parallel. This approach has been taken by
Scarpazza et al.~\cite{ScarpazzaVP07} with a DFA-based string matching
system for network security on the IBM Cell BE processor.
Similarly, Wang et
al.~\cite{WangHL10} investigated parallel architectures for packet
inspection based on DFAs. Both approaches assume multiple input streams
and a vast number of
patterns (i.e., virus signatures), which is common with network security
applications. However, neither
approach parallelizes the DFA membership algorithm itself, which is
required to improve
applications with single, long-running membership tests such as DNA
sequence analysis.

Scarpazza et al.\ utilize the SIMD units of the Cell BE's synergistic
processing units to match multiple input streams in parallel. However,
their vectorized DFA matching
algorithm contains several SISD instructions and the reported speedup
from 16-way vectorization is only a factor of 2.51.

Recent research efforts focused on speculative computations to
parallelize DFA membership tests.
Holub and \v{S}tekr~\cite{Holub:2009} were the first to split the input
string into chunks and distribute chunks among available processors. Their
speculation introduces a substantial amount of redundant computation, which
restricts the obtainable speedup for general DFAs
to~$\mathcal{O}(\frac{\lvert P\rvert}{\lvert Q\rvert})$, where
$\lvert P\rvert$ is
the number of processors, and $\lvert Q\rvert$ is the number of DFA
states.
Their algorithm degenerates to a speed-down when
$|\States|$ exceeds the number of processors.
To overcome this problem, Holub and \v{S}tekr specialized their
algorithm for
$k$-local DFAs. A DFA is $k$-local if for every word of length~$k$ and
for all states~$p,q\in Q$ it holds that
$\TFE(p, w) = \TFE(q,w)$. Starting the matching operation $k$ symbols
ahead of a given
chunk will synchronize the DFA into the correct initial state by
the time matching reaches the beginning of the chunk, which eliminates
all speculative computation. Holub and  \v{S}tekr achieve
a linear speedup of $\mathcal{O}(\lvert P\rvert)$ for $k$-local automata.

Jones et al.~\cite{Jones2009} reported that with the IE~8 and Firefox
web browsers
3--40\% of the execution-time is spent parsing HTML documents. To speed
up browsing, Jones et al.\
employ speculation to parallelize token detection (lexing) of HTML
language front-ends.
Similar to Holub and \v{S}tekr's $k$-local automata, they use the
preceding~$k$
characters of a chunk to synchronize a DFA to a particular state.
Unlike $k$-locality,
which is a static DFA property, Jones et al.\
speculate the DFA to be in a particular, frequently occurring DFA state
at the beginning of a chunk. Speculation fails if the DFA turns
out to be in a different state, in which case the chunk
needs to be re-matched. Lexing HTML documents results in frequent
matches, and the structure of regular expressions is reported to be
simpler than, e.g., virus signatures~\cite{Luchaup2011}. Speculation is
facilitated by the fact that the state at the beginning
of a token is always the same, regardless where lexing started. A
prototype implementation is reported to scale up to six of the eight
synergistic processing units of the Cell BE.

The speculative parallel pattern matching (SPPM) approach by
Luchaup et al.~\cite{Luchaup2011,Luchaup2009} uses speculation to match
the increasing network line-speeds
faced by intrusion prevention systems. SPPM DFAs represent virus
signatures. Like Jones et al.,
DFAs are speculated to be in a particular, frequently occurring DFA state
at the beginning of a chunk. SPPM starts the speculative matching
at the beginning of each chunk. With every input character, a
speculative matching process stores the encountered DFA
state for subsequent reference. Speculation fails if the DFA turns out
to be in a different state at the beginning of a speculatively matched
chunk. In this case, re-matching continues until the DFA
synchronizes with the saved history state (in the worst case, the
whole chunk needs to be re-matched). A single-threaded SPPM
version is proposed to improve performance by issuing multiple
independent memory accesses in parallel.
Such pipelining (or interleaving) of DFA matches is orthogonal to
our approach, which
focuses on latency rather than throughput.

SPPM assumes all regular expressions to be suffix-closed, which is the
common scenario with intrusion prevention systems;
A regular expression is suffix-closed if matching a given string~$w$
implies that $w$ followed by any suffix is matched, too. A suffix-closed
regular language has the property that
$x\in L \Leftrightarrow\forall w \in\Sigma^*: xw\in L$.

The speculative parallel DFA membership test reported by
Ko et al.~\cite{Ko2014} is to parallelize DFA membership tests
for multicore, SIMD, and cloud computing environments.
This technique is one of the speculative parallel matching methods
by searching arbitrary regular expressions. It requires dividing
the input string into chunks, matching chunks in parallel, and combining
the matching results. When the input string is partitioned, the
algorithm decides the amount of characters of each chunk depending on
the number of possible start states.
Unlike the previous approaches of parallel
membership tests, it is {\em failure-free\/}, which means speed-downs
never happen by maintaining the sequential semantics.

Another approach of parallel pattern matching was investigated by
Mytkowicz et al.~\cite{Mytkowicz2014}. It provides a parallel
algorithm for DFAs, which breaks the dependencies of state transitions
by enumerating computations from all possible start states on each
input character. With the enumeration algorithm, the number of
state transitions decreases even though all states are considered as
possible start states at the beginning of matching.

To remove the overhead from the speculative parallel pattern
matching, a simultaneous finite automaton is introduced
by Sin'ya et al.~\cite{Sinya2013}. Because the main problem of parallel
pattern matching is the dependency of state transitions, it extends
an automaton so that it involves the simulation of state transitions.
Although it can remove the dependency, extending an automaton to a
SFA is costly because the algorithm is based on the well-known subset
construction algorithm, which requires considerable time.
And our approach makes their algorithm efficiently by parallelizing it.

One of our optimization methods is hashing and we utilized hashing to accomplish faster 
execution. Stern et al.~\cite{Stern95improvedprobabilistic} applied hashing to save memory
because the verification of complex protocols suffered from the state explosion problem.
Even though we have the same state explosion problem, we approached it in the non-probabilistic
way whereas they considered the probabilistic method by compacting states in the hash table.
                                                
\section{Conclusions}
\label{sec:conclusions}
Although string pattern matching with FAs has been widely used in various
areas, parallelizing its sequential algorithm was difficult because of the 
dependency of state transitions. As a solution to break this, SFAs
were introduced. Although SFA matching achieved 
a lot of speedups, constructing SFAs was costly in terms of  
execution time because the algorithm was implemented sequentially and 
required many number of computations.
Also, the size of an SFA is significantly bigger than that of an original DFA.

In this paper, we presented several optimization methods for 
the SFA construction algorithm on multicore architecture platforms.
We adapted fingerprints method provided by Rabin with the help of Barrett
reduction algorithm and AES, SSE instruction set to reduce the number of comparisons of
SFA states because a comparison requires considerable time.
Hashing was introduced to support fingerprints for further
optimizations and it derived the $\mathcal{O}(1)$ time complexity of the algorithm in most 
cases. Our proposed algorithm is nonblocking and exploits various
parallelism sources after investigation; instruction-level, data-level, and task-level
parallelism with coarse, medium, and fine granularities. Thread level parallelization is used for each parallelism sources; instruction, data and task-level parallelism.
We also suggested the static distribution calculates appropriate amount for each thread in advance so every thread
can have and process almost same amount of work simultaneously. 
And transposing the transition table was implemented in consideration of the locality of memory accesses.
As a result, our overall speedups over the sequential baseline which we could observe on our servers are up to 118541x on the AMD system and 2113968x on the Intel system.





\bibliographystyle{IEEEtran}
%

\bibliography{SDFA}



\end{document}